\documentclass[10pt]{article}
\usepackage{amsmath}
\usepackage{amssymb}

\newcommand{\de}{\partial}
\newcommand{\lam}{\lambda}
\newcommand{\tet}{\theta}
\newcommand{\dif}{\,\text{d}}

\begin{document}
\title{\Large\bf Compatibility of radial, Lorenz and harmonic gauges}

\author{Elena Magliaro{\it ${}^{ac}$}, Claudio Perini{\it ${}^{bc}$}, Carlo Rovelli{\it ${}^{c}$}\\[1mm]
\small\it ${}^a$Dipartimento di Fisica, Universit\`a degli Studi Roma Tre, I-00146 Roma, EU\\[-1mm]
\small\it ${}^b$Dipartimento di Matematica, Universit\`a degli Studi Roma Tre, I-00146 Roma, EU\\[-1mm]
\small\it ${}^c$Centre de Physique Th\'eorique de Luminy%
\footnote{Unit\'e mixte de recherche (UMR 6207) du CNRS et des Universit\'es
de Provence (Aix-Marseille I), de la M\'editerran\'ee (Aix-Marseille II) et du Sud (Toulon-Var); laboratoire affili\'e \`a la FRUMAM (FR 2291).}, Case 907, F-13288 Marseille, EU}
\date{\small\today}
\maketitle

\begin{abstract}\noindent
We observe that the radial gauge can be consistently imposed \emph{together} with the Lorenz gauge in Maxwell theory, and with the harmonic traceless gauge in linearized general relativity. This simple observation has relevance for some recent developments in quantum gravity where
the radial gauge is implicitly utilized. 
\end{abstract}
\vskip1cm

\section{Introduction}
\vskip5mm
The radial gauge, or Fock-Schwinger gauge \cite{Fock,Schwinger}, is defined by 
\begin{equation}\label{Max}
x^\mu A_\mu=0
\end{equation}
in Maxwell theory, and by 
\begin{equation}\label{GR}
x^\mu h_{\mu\nu} =0 
\end{equation}
in linearized general relativity. Here $x\!=\!(x^\mu)$ are Lorentzian (or Euclidean) spacetime coordinates in $d+1$ spacetime dimensions, where  \,$\mu\!=\!0,1,\dots,d$\,;\,\,$A_\mu(x)$\, is the electromagnetic potential, and $h_{\mu\nu}(x)=g_{\mu\nu}(x)-\eta_{\mu\nu}$ is the perturbation of the metric field $g_{\mu\nu}(x)$ around the background Minkowski (or Euclidean) metric $\eta_{\mu\nu}$ used to lower and raise indices.  The radial gauge has been considered with various motivations. For instance, radial--gauge perturbation theory was studied in 
\cite{Modanese,Menotti-Modanese,Leupold,Leupold2}, where an expression for the propagator and Feynman rules in this gauge were  derived.  A number of recent papers implicitly use this gauge in the context of nonperturbative Euclidean loop quantum gravity  \cite{QGbook,Propagator,Propagator2,Speziale,LivSpez}.   These papers, indeed, consider a spherical region in 4d Euclidean spacetime, and identify the degrees of freedom on the 3d boundary $\Sigma$ of this region with the degrees of freedom described by Hamiltonian loop quantum gravity. The last is defined in a ``temporal'' gauge where the field components in the direction {\em normal to the boundary surface} $\Sigma$ are gauge fixed. Since the direction normal to a sphere is radial, this procedure is equivalent to imposing the radial gauge (\ref{GR}) in the linearization around flat spacetime.

The radial gauge is usually viewed as an \emph{alternative} to the commonly used Lorenz and harmonic gauges, defined respectively by   
\begin{equation}\label{lorenz} 
\partial_\mu A^\mu=0
\end{equation}
in Maxwell theory and by 
\begin{equation}\label{ht}
\partial_\mu h^\mu{}_\nu-\frac12\, \partial_\nu h^\mu{}_\mu=0
\end{equation} 
in linearized general relativity.  Here we observe, instead, that the radial gauge is \emph{compatible} with the Lorenz and the harmonic gauges. That is, if $A_\mu$ and $h_{\mu\nu}$ solve the Maxwell and the linearized Einstein equations, then they can be gauge-transformed to fields $A'_\mu$ and $h'_{\mu\nu}$ satisfying (\ref{Max},\ref{GR}) \emph{and} (\ref{lorenz},\ref{ht}).  This is analogous to the well known fact (see for instance   \cite{Wald}) that the Lorenz and the harmonic gauges can be imposed simultaneously with the temporal gauge
\begin{subequations}
\begin{align}\label{temporal}
&A_0=0\;,\\&\label{temporal_gr}h_{0\mu}=0\;.
\end{align}
\end{subequations}

This observation is of interest in relation to the quantum gravity results mentioned above, for the following reason.  One of the aims of these calculations is to compute the $n$-point functions from loop quantum gravity and compare them with the corresponding expressions obtained in the conventional perturbative expansion of quantum general relativity.  Agreement at large distance could then be taken as evidence that the nonperturbative quantum theory has the correct low energy limit; while the differences at short distance reflect the improved ultraviolet behavior of the nonperturbative theory.  Difficulties for the comparison could be expected, however, due to the fact that, as observed, the nonperturbative calculations are in the radial gauge, while the  perturbative quantum theory is mostly known in the harmonic gauge \cite{Dewitt,review}. The observation in this paper shows that this is not a problem, since the two gauges are compatible. The $n$-point functions computed from loop quantum gravity can be directly compared with the appropriate components of those derived in the harmonic gauge.

We find convenient, below, to utilize the language of general covariant tensor calculus. To avoid confusion, let us point out that this does not mean that we work on a curved spacetime.  We are only concerned here with Maxwell theory on flat space and with linearized general relativity also on flat space.  Tensor calculus is used below only as a tool for dealing in compact form with expressions in the hyperspherical coordinates that simplify the analysis of the radial gauge. 

Maxwell theory is discussed in Section 2.  Gravity is discussed in section 3.   We work in an arbitrary number of dimensions, and we cover  the Euclidean and the Lorentzian signatures at the same time. That is, we can take either \,$(\eta_{\mu\nu})\!=\!\text{diag}[1,1,1,1, ...]$\, or \,$(\eta_{\mu\nu})\!=\!\text{diag}[1,-1,-1,-1,...]$\,. The analysis is local in spacetime and disregards singular points such as the origin.

\section{Maxwell theory}\label{em}
\vskip5mm
In this section we show the compatibility between Lorenz and radial gauge in electromagnetism. 
Maxwell vacuum equations are 
\begin{equation}
	\de_\nu F^{\nu\mu}=0\,, 
\end{equation}
where $F_{\mu\nu}\!=\de_\mu A_\nu-\de_\nu A_\mu$\,. That is
\begin{equation}
\label{maxwell}
\Box A_\mu-\de_\mu\de_\nu A^\nu=0\;, 
\end{equation}
where \,$\Box=\eta^{\mu\nu}\de_\mu \de_\nu$\,. 
This equation is of course invariant under the gauge transformation
\begin{equation}\label{gauge}
	A_\mu\rightarrow A'_\mu=A_\mu+\de_\mu\lambda\;. 
\end{equation} 

\subsection{Temporal and Lorenz gauge}\label{tl}
\vskip5mm
We begin by recalling how one can derive the well-know result that the Lorenz and \emph{temporal} gauges are compatible.  This is a demonstration that can be found in most elementary books on electromagnetism; we recall it here in a form that we shall reproduce below for the radial gauge.

Let us write \,$(x^\mu)\!=\!(x^0,x^i)\!=\!(t,\vec x\,)$\,, where \,$i\!=\!1,\ldots,d$\,.  Let $A_\mu$ satisfy the Maxwell equations \eqref{maxwell}. We now show that there is a gauge equivalent field $A'_\mu$ satisfying the temporal as well as the Lorenz gauge conditions. That is, there exist a scalar function $\lambda$ such that $A_\mu'$ defined in \eqref{gauge} satisfies \eqref{temporal} and \eqref{lorenz}. The equation \eqref{temporal} for $A_\mu'$ defined in \eqref{gauge} gives 
$A_0+\partial_0\lambda=0$\,, with the general solution 
 \begin{equation}
 \label{emt_soltemp}
 \lambda(t,\vec x\,)=-\int_{t_0}^t\! A_0(\tau,\vec x\,)\dif\tau+\tilde\lam(\vec x\,)\; ,
 \end{equation}
where $\tilde\lam(\vec x\,)$ is an integration ``constant'', which is an arbitrary function on
the surface $\Sigma$ defined by \,$t\!=\!t_0$\,.   
Can $\tilde\lam(\vec x\,)$ (which is a function of $d$ variables) be chosen in such a way that  the Lorenz gauge condition (which is a function of $d+1$ variables) is satisfied?  To show that this is the case, let us first fix  $\tilde\lam(\vec x\,)$ in such a way that  the Lorenz gauge condition is satisfied {\em on} $\Sigma$\,.  Inserting $A'_\mu$ in \eqref{lorenz} and using \eqref{temporal} we have 
 \begin{equation}
\partial_\mu A'^\mu=\partial_i A'^i=\partial_i A^i+\Delta\lambda=0\,,
 \end{equation}
where $\Delta = \partial_i\partial^i$ is the Laplace operator\footnote{Minus the Laplace operator in the Lorentzian case.} on $\Sigma$\,. The restriction of this equation to $\Sigma$ gives the Poisson equation
  \begin{equation}
\Delta\tilde\lambda(\vec x\,)=-\partial_i A^i(t_0,\vec x\,)\;,
 \end{equation}
which determines  $\tilde \lambda(\vec x\,)$\,. With $\tilde\lambda(\vec x\,)$ satisfying this equation, $A'_\mu$ satisfies the temporal gauge condition everywhere and the Lorenz gauge condition on $\Sigma$\,.  However, this implies immediately that $A'_\mu$ satisfies the Lorenz gauge condition everywhere as well, thanks to the Maxwell equations. In fact, the time component of \eqref{maxwell} reads
\begin{equation}
	\Box A'_0-\de_0\de_\nu A'^\nu=-\de_0(\de_\nu A'^\nu)=0\;.
\end{equation}
That is: for a field in the temporal gauge, the Maxwell equations  imply that if the Lorenz gauge is satisfied on $\Sigma$ then it is satisfied everywhere. 
%%%%%%%%%%%%%%%%%%%%%%%%%%%%%%%%%%%%%%%%%%%%%%%%%%%%%%%%
\subsection{Radial and Lorenz gauge}
\vskip5mm
We now show that the {\em radial} and Lorenz gauge are compatible, following steps similar to the ones above.  We want to show that there exists a function \,$\lambda$\, such that \,$A'_\mu$\, defined in \eqref{gauge} satisfies \eqref{Max} and \eqref{lorenz}, assuming that \,$A_\mu$\, satisfies the Maxwell equations.  

Due to the symmetry of the problem, it is convenient to use polar coordinates. We write these as \,$(x^a)\!=\!(x^r,x^i)\!=\!(r,\vec x\,)$\,, where \,$r\!=\sqrt{|\eta_{\mu\nu}x^\mu x^\nu|}$\, is the $(d+1)$-dimensional radius and ${\vec x}=(x^i)$ are three angular coordinates.  In these coordinates the metric tensor \,$\eta_{\mu\nu}$\, takes the simple form
\begin{equation}\label{emr_ds}
\dif s^2= \gamma_{ab}(r,\vec x\,)\dif x^a\dif x^b= \dif r^2+r^2 \xi_{ij}(\vec x\,)\dif x^i\dif x^j\;,
\end{equation}
where $\xi_{ij}(\vec x\,)$  is independent from $r$ and is the metric of a 3-sphere of unit radius in the Euclidean case, and the metric of an hyperboloid of unit radius in the Lorentzian case. It is easy to see that in these coordinates, the radial gauge condition \eqref{Max} takes the simple form 
\begin{equation}\label{Ar}
A'_r=0\;.
\end{equation}
Inserting the definition of $A'_\mu$ 
gives 
\begin{equation}
\label{emr_radial}
\partial_r\lam=-A_r\;,
\end{equation}
with the general solution
\begin{align}
	\lambda(r,\vec x\,)=-\int_{r_0}^r\!\!A_r(\rho,\vec x\,) \dif \rho+\tilde\lambda(\vec x\,)\;,
\end{align}
where the integration constant $\tilde\lambda$ is now a function on the surface $\Sigma$ defined by \,$r\!=\!r_0$\,. The surface $\Sigma$ is a $d$-sphere in the Euclidean case and a $d$-dimensional hyperboloid in the  Lorentzian case.  As in the previous section, we fix $\tilde\lambda(\vec x\,)$ by requiring the Lorenz condition to be satisfied on $\Sigma$\,.  It is convenient to use general covariant tensor calculus in order to simplify the expressions in polar coordinates.  In arbitrary coordinates, the Lorenz condition reads 
\begin{equation}
\label{emr_Lorenz}
\nabla_a A'^a=\frac{1}{\sqrt{\gamma}}\partial_a \big(\sqrt{\gamma} A'^a\big)=0\;,
\end{equation}
where $\nabla_a$ is the covariant derivative, $A_b\!=\!A^a g_{ab}$\,, and $\gamma$ is the determinant of $\gamma_{ab}$\,. This determinant has the form \,$\gamma\!=\!r^{2d}\xi$\,, where $\xi$ is the determinant of $\xi_{ij}$\,. When the radial gauge is satisfied, \eqref{emr_Lorenz} reduces to
\begin{equation}\label{rl}
\partial_i \big(\sqrt{\xi} A'^i\big)=0\;.
\end{equation}
Let us now require that $A'_\mu$ satisfies this equation on $\Sigma$\,.  Using \eqref{gauge}, this
requirement  fixes $\tilde\lambda$ to be the solution of a Poisson equation on $\Sigma$\,, that is 
\begin{equation}
\label{aa} 
\Delta \tilde\lambda =-\frac{1}{\sqrt{\xi}} \partial_i \big(\sqrt{\xi} A^i\big)\;,
\end{equation}
where the Laplace operator is \,$\Delta=\nabla_i\,\xi^{ij}\nabla_j$\,. 
In arbitrary coordinates, Maxwell equations read
\begin{align}\label{emr_Maxwell}
	\nabla_a F^{ab}=\frac{1}{\sqrt\gamma}\de_a\big(\sqrt\gamma \,F^{ab}\big)=0\;,
\end{align}
where
\begin{align}
	F^{ab}=\nabla^a A^b-\nabla^b A^a\,.
\end{align}
Consider the radial ($b=r$) component of \eqref{emr_Maxwell}; since $A'_r\!=\!0$\,, using the form \eqref{emr_ds} of the metric, we have
\begin{align}\label{emr_Maxr}\nonumber
			&\frac{1}{\sqrt{\gamma}}\de_a(\sqrt\gamma F^{a r})=\frac{1}{\sqrt{\gamma}}\de_a(\sqrt\gamma \gamma^{ab}F_{b r})
			=\frac{1}{\sqrt{\gamma}}\de_a(\sqrt{\gamma}\gamma^{ab}(\de_b A'_r-\de_r A'_b))=\\
	    &=-\frac{1}{\sqrt{\xi}}\de_i\Big(\sqrt{\xi}\,\frac{\xi^{ij}}{r^2}\de_r A'_j\Big)=
	    -\frac{1}{r^2\sqrt{\xi}}\de_r\de_i(\sqrt{\xi}\xi^{ij}A'_j)=0\;,
\end{align}
which shows that the Lorenz gauge condition \eqref{rl} is satisfied everywhere if it satisfied on $\Sigma$\,.  This shows that we can find a function $\lambda$ such that both the radial and the Lorenz gauge are satisfied everywhere. 
%%%%%%%%%%%%%%%%%%%%%%%%%%%%%%%%%%%%%%%%%%%%%%%%%%%%%%%%%%%%%%%%%%%%%%%%%%%
\section{Linearized general relativity}\label{gr}
\vskip5mm
We now consider the compatibility between the radial gauge and the harmonic traceless gauge (also known as transverse traceless gauge\cite{Wald}) in linearized general relativity.  Einstein equations in vacuum are given by the vanishing of the Ricci tensor. If \,$|h_{\mu\nu}(x)|\ll 1$\,, and we linearize these equations in $h_{\mu\nu}$\,, we obtain the 
linearized Einstein equations 
\begin{equation}\label{einstein}
\de_\mu\de_\nu h^\alpha_{\;\;\alpha}+\de_\alpha\de^\alpha h_{\mu\nu}-\de_\mu\de^\alpha h_{\alpha\nu}-\de_\nu\de^\alpha h_{\alpha\mu}=0\;.
\end{equation}
Under infinitesimal coordinate transformations, 
\begin{equation} \label{gaugegr}
h_{\mu\nu}\rightarrow h'_{\mu\nu}=h_{\mu\nu}+\frac{1}{2}(\de_\mu\lam_\nu+\de_\nu\lam_\mu)\;,
\end{equation}
where the factor $1/2$ is inserted for convenience. These are gauge transformations of the linearized theory.  The harmonic gauge is defined by the condition
\begin{equation}\label{grintro_harmonic}
\nabla^\nu \nabla_\nu x^\mu=0\;,
\end{equation}
where $\nabla_\nu$ is the covariant partial derivative\footnote{Notice that \eqref{grintro_harmonic} means the covariant Laplacian of $d\!+\!1$ scalars ($d\!+\!1$ coordinates), not the covariant Laplacian of a $(d\!+\!1)$-vector.}; in the linearized theory  \eqref{grintro_harmonic} reduces to
\begin{equation}\label{harmonic}
\de_\nu h^{\nu\mu}-\frac{1}{2}\de^\mu h^\nu_{\;\;\nu}=0\;,
\end{equation}
and in this gauge the Einstein equations \eqref{einstein} read simply
\begin{align}
	\Box h_{\mu\nu}=0\;.
\end{align}
%%%%%%%%%%%%%%%%%%%%%%%%%%%%%%%%%%%%%%%%%%%%%%%%%%%%%%%%%%%%%%%%%%%%%%%%%%%%%%%%%%%%
\subsection{Temporal and harmonic gauge}
\vskip5mm
As we did for Maxwell theory, we begin by recalling how the compatibility between \emph{temporal} and harmonic gauge can be proved. Start by searching a gauge parameter $\lambda_\mu$ that takes $h_{\mu\nu}$ to the temporal gauge $h'_{0\nu}=0$\,. Equation \eqref{temporal_gr} gives
\begin{align}
h_{0\mu}+\frac{1}{2}(\de_0\lam_\mu+\de_\mu\lam_0)=0 
\end{align}
with the general solution 
\begin{subequations}
\begin{align}\label{grt_soltemp1}
	&\lam_0(t,\vec x\,)=-\int_{t_0}^t\! h_{00}(\tau,\vec x\,)\dif \tau+\tilde\lam_0(\vec x\,)\;,\\\label{grt_soltempi2}
	&\lam_i(t,\vec x\,)=-\int_{t_0}^t\!\Big(2h_{0i}(\tau,\vec x\,)+\de_i \lambda_0(\tau, \vec x\,)\Big)\dif \tau+\tilde\lam_i(\vec x\,)\;, 
\end{align}
\end{subequations}
where the integration constants $\tilde\lam_\mu(\vec x\,)$ are functions on the 3d surface $\Sigma$ defined by \,$t=t_0$\,. 
 Next, we fix $\tilde\lambda_i$ by imposing the harmonic gauge condition \eqref{harmonic} on $\Sigma$\,. Since we are in temporal gauge, this gives
\begin{align}
\Delta \tilde\lambda_j=-2\de_i h^i_{\;\,j}+\partial_j h^i_{\;\,i}\,,
\end{align}
which can be clearly solved on $\Sigma$\,. The time-time component of Einstein equations becomes
\begin{align}\label{grt_00}
	\de^2_t h'^i_{\;\;\,i}=0\;,
\end{align}
whose only well behaved solution is $h'^i_{\;\;i}=0$\,; so in the temporal gauge the invariant trace of $h'_{\mu\nu}$ vanishes:
\begin{align}\label{traceless}
	h'^\mu_{\;\;\,\mu}=\eta^{\mu\nu}h'_{\mu\nu}=0\;,
\end{align}
and the harmonic condition \eqref{harmonic} takes the simpler form
\begin{align}
	\de_\nu h'^{\nu\mu}=0\;,
\end{align}
similar to the Lorenz gauge. Now the $(t,i)$ components of Einstein equations read
\begin{align}
	\de_t\de_j h'^j_{\;\;\,i}=0\;,
\end{align}
which give \,$\de_j h'^j_{\;\;\,i}=0$\, everywhere, once imposed on $\Sigma$\,.

\subsection{Radial and harmonic gauge}\label{subs_radialharmonic}
\vskip5mm
Let us finally come to the compatibility between the \emph{radial} and harmonic gauges. 
We return to the polar coordinates used in the Maxwell case.  In these coordinates, the radial gauge condition \eqref{GR} reads 
\begin{equation}
h'_{rr}=h'_{ri}=0\;.
\end{equation} 
Inserting the gauge transformation \eqref{gaugegr} gives
\begin{subequations}
\begin{align}
&\de_r\lambda_r=-h_{rr}\;,\\
&\de_r\lambda_i+\de_i\lambda_r-\frac{2}{r}\lambda_i=-2 h_{ri}\;,
\end{align}
\end{subequations}
with the general solution
\begin{subequations}
\begin{align}
	&\lam_r(r,\vec x\,)=-\int_{r_0}^r h_{rr}(\rho,\vec x\,)\dif \rho+\tilde\lam_r(\vec x\,)\;,\\
	&\lam_i(r,\vec x\,)=-r^2\!\!\int_{r_0}^r\!\frac{2h_{ri}(\rho,\vec x\,)+\de_i\lam_r(\rho,\vec x\,)}{\rho^2}\dif \rho
	+r^2 \ \tilde\lambda_i(\vec x\,)\; , 
\end{align}
\end{subequations}
where $\tilde\lam_r,\tilde\lambda_i$ are functions on the surface $\Sigma$ given by $r=r_0$. 
We can then fix $\tilde\lambda_i$ by imposing the harmonic condition on $\Sigma$ precisely as before. In the polar coordinates \eqref{emr_ds}, we have easily the following rules for the Christoffel symbols:
\begin{align}\label{grr_Christrules}
	\Gamma^a_{\;\;rr}=0\;,\;\;\;\;\; \Gamma^i_{\;\;jr}=\frac{1}{r}\delta^i_{\;\;j}\;,\;\;\;\;\; \Gamma^r_{\;\;ra}=0\;.
\end{align} 
We note also that $\Gamma^i_{\;jk}$ is independent of $r$\,. Consider the $(r,r)$ component of Einstein equations:
\begin{align}
	\nabla_r\nabla_r h'^{a}_{\,\;\;a}+\nabla_{a}\nabla^{a} h'_{rr}-\nabla_r\nabla^{a} h_{a r}-\nabla_r\nabla^a h'_{a r}=0\;.
\end{align}
Taking into account \eqref{emr_ds} and \eqref{grr_Christrules}, it is verified after a little algebra that the previous equation becomes
\begin{align}
\de_r^2 h'^{a}_{\;\;\,a}+\frac{2}{r}\de_r h'^{a}_{\;\;\,a}\label{grr_00}=0\;,
\end{align}
which is a differential equation for the trace \,$h'^a_{\;\;\,a}$\,. Its only solution well-behaved at the origin and at infinity is \,$h'^a_{\;\;\,a}\!=0$\,. Using this, the $(r,i)$ components of Einstein equations read:
\begin{align}\label{grr_Einstrmu}
	\nabla_a\nabla^a h'_{ri}-\nabla_r\nabla^a h'_{a i}-\nabla_i\nabla^a h'_{a r}
	=-\de_r \nabla_a h'^a_{\;\;\;i}=0\;,
\end{align}
and the harmonic condition is simply
\begin{align}\label{grr_harmonic}
	\nabla_a h'^{ab}=0\;.
\end{align}
Equation \eqref{grr_Einstrmu} shows immediately that the $b\!=\!i$ components of the gauge condition \eqref{grr_harmonic} hold everywhere if they hold on $\Sigma$\,. The vanishing of the $b\!=\!r$ component of \eqref{grr_harmonic} follows immediately since, using \eqref{grr_Christrules}, we have
\begin{align}
\nabla_a h'^a_{\;\;\,r}=-\frac{1}{r}h'^a_{\;\;\,a}=0\;.
\end{align}
Therefore the harmonic gauge condition, the radial gauge condition and the vanishing of the trace are all consistent with one another.
%%%%%%%%%%%%%%%%%%%%%%%%%%%%%%%%%%%%%%%%%%%%%%%%%%%%%%%
\appendix
\section*{Appendix}
\vskip5mm
We give here for convenience the definition of the coordinate systems in four dimensions we used in this work, followed by the respective line elements (metrics).
\paragraph{\textnormal{\emph{Polar hyperspherical coordinates}}}
\begin{subequations}
\begin{align}
	&x^0=r \cos\psi\\
	&x^1=r \sin\theta\cos\phi\sin\psi\\
	&x^2=r \sin\theta\sin\phi\sin\psi\\
	&x^3=r \cos\theta\sin\psi
\end{align}
\end{subequations}
Euclidean line element:
\begin{align}
	\dif s^2=\dif r^2+r^2(\sin^2\psi\dif\tet^2+\sin^2\tet\sin^2\psi\dif\phi^2+\dif\psi^2)
\end{align}
\paragraph{\textnormal{\emph{Polar hyperbolic coordinates (future light cone)}}}
\begin{subequations}
\begin{align}
  &x^0=\rho\,\cosh\chi\\
	&x^1=\rho\,\sin\theta\cos\phi\sinh\chi\\
	&x^2=\rho\,\sin\theta\sin\phi\sinh\chi\\
	&x^3=\rho\,\cos\theta\sinh\chi
	\end{align}
\end{subequations}
Minkowski line element:
\begin{align}
	\dif s^2=\dif \rho^2-\rho^2(\sinh^2\chi\dif\tet^2+\sin^2\tet\sinh^2\chi\dif\phi^2+\dif\chi^2)
\end{align}


\begin{thebibliography}{99}
\bibitem{Fock} V. A. Fock, ``Proper time in classical and quantum mechanics'', Sov. Phys. 12 (1937) 404
\bibitem{Schwinger} J. Schwinger, ``On gauge invariance and vacuum polarization'', Phys. Rev. 82 (1952) 684
\bibitem{Modanese} G. Modanese, ``The propagator in the radial gauge'', J. Math. Phys. 33 (1992) 1523 
\bibitem{Menotti-Modanese} P. Menotti, G. Modanese, D. Seminara, ``The radial gauge propagators in quantum gravity'', Ann. Phys. 224 (1993) 110 [arXiv:hep-th/9209028]
\bibitem{Leupold} S. Leupold, H. Weigert, ``Radial propagators and Wilson loops'', Phys. Rev. D 54 (1996) 7695 [arXiv:hep-th/9604015]
\bibitem{Leupold2} S. Leupold, ``Feynman rules in radial gauge'' (1996) [arXiv:hep-th/9609222]
\bibitem{QGbook} C. Rovelli, \emph{Quantum Gravity}, Cambridge University Press (Cambridge, 2004)
\bibitem{Propagator} C. Rovelli, ``Graviton propagator from background-independent quantum gravity'', Phys. Rev. Lett. 97 (2006) 151301 [arXiv:gr-qc/0508124]
\bibitem{Propagator2} E. Bianchi, L. Modesto, C. Rovelli, S. Speziale, ``Graviton propagator in loop quantum gravity'', Class. Quant. Grav. 23 (2006) 6989 [arXiv:gr-qc/0604044]
\bibitem{Speziale} S. Speziale, ``Towards the graviton from spinfoams: The 3d toy model'', JHEP 05 (2006) 039 [arXiv:gr-qc/0512102]
\bibitem{LivSpez} E. R. Livine, S. Speziale, ``Group integral techniques for the spinfoam graviton propagator'', JHEP 11 (2006) 092 [arXiv:gr-qc/0608131]
\bibitem{Wald} R. M. Wald, \emph{General relativity}, University Of Chicago Press (Chicago, 1984)
\bibitem{Dewitt} B. S. DeWitt, `` Quantum Theory of Gravity. III. Applications of the Covariant Theory'', Phys. Rev. 171 (1968) 1834
\bibitem{review} A. Akhundov, A. Shiekh, ``A Review of Leading Quantum Gravitational Corrections to Newtonian Gravity'' (2006) [arXiv:gr-qc/0611091]
\end{thebibliography}
\end{document}